\documentstyle[12pt,aasms4]{article}
\makeatletter
\def\jnl@aj{AJ}
\ifx\revtex@jnl\jnl@aj\let\tablebreak=\nl\fi
\makeatother

\def \etal {et al.\thinspace}

\def \teff {$T_{eff}$}

\begin{document}

\title{Resonance Averaged Photoionization Cross Sections for Astrophysical 
Models}
\author{Manuel A. Bautista}
\affil{Laboratory for High Energy Astrophysics, Code 662}
\affil{NASA Goddard Space Flight Center, Greenbelt, MD\ 20771}
\author{Patrizia Romano, \& Anil K. Pradhan} 
\affil{Department of Astronomy, The Ohio State University}
\affil{174 West 18th Avenue, Columbus, OH\ 43210-1106}

\begin{abstract}

We present ground state photoionization cross       
sections of atoms and ions averaged over resonance 
structures for photoionization modeling of astrophysical sources.
The detailed cross sections calculated in the close-coupling
approximation using the R-matrix method, with resonances delineated at 
thousands of energies,  are taken from
the Opacity Project database TOPbase and the Iron Project,
including new data for the low ionization stages of iron Fe~I -- V.
The resonance-averaged cross sections are obtained by convolving the detailed 
cross sections with a 
Gaussian distribution over the autoionizing resonances.
This procedure is expected to minimize errors in the derived ionization rates 
that could result from small uncertainties in computed positions of resonances, 
while preserving the overall resonant contribution to the  
cross sections in the important near threshold regions. 
The detailed photoionization cross sections at low photon energies
are complemented by new relativistic 
distorted-wave calculations for Z$\le 12$, and from central-field 
calculations for Z$> 12$ at high energies, including inner-shell
ionization.
The effective cross sections are then represented 
by a small number of points that can be readily interpolated linearly for
practical applications; a Fortran subroutine and data are available. 
The present numerically averaged cross sections
are compared with analytic fits that do not accurately represent the
effective cross sections in regions dominated by resonances.

\end{abstract}



\section{Introduction}

Photoionization cross sections are necessary for the computation of 
photoionization and recombination rates 
for ionization balance in astrophysical plasmas (e.g. Kallman and
Krolik 1991, Shull \& Van Steenberg 1982, Sutherland \& Dopita 1993). 
Accurate cross sections have been calculated in the close-coupling
approximation using the R-matrix method,
for most astrophysically important atoms and ions under the Opacity 
Project (OP; Seaton \etal 1994) and the Iron Project (IP; Hummer \etal 1993).
The cross sections incorporate, in an {\it ab initio} manner,
the complex autoionizing resonance structures 
that can make important contributions to total photoionization
rates. These data are currently available from the electronic  database
of the Opacity Project TOPbase 
(Cunto \etal 1993; {\it The Opacity Project Team} 1995). 

 Resonant phenomena have been shown to be of crucial, often dominant, 
importance in electron-ion scattering, photoionization, and
recombination processes (see, for example, the references for the OP and
the IP). Most of the related calculations for such
atomic processes have been carried out in the close-coupling (hereafter
CC)
approximation that quantum mechanically couples the open and closed 
channels responsible for the continuum, and 
the quasi-bound resonant states, respectively.
Photoionization calculations in the CC
approximation consist of an expansion over the states of the
(e + ion) system, with a number of excited states of the residual (photoionized)
ion (also called the ``target"). The CC approximation thereby
 includes, in an {\it ab initio} manner, several
infinite Rydberg series of resonances converging on to the
states of the target ion. The resonances are
particularly prominent in the near-threshold region
due to strong electron correlation effects, 
and the accuracy of the calculated cross section depends 
on the representation of the resonances. 
Furthermore, singular resonant features may sometimes dominate the cross
section over an extended energy region.  A prime example of such
resonances, in addition to the Rydberg series of resonances and
the broad near-threshold resonances, are the so called
``photoexcitation-of-core" (PEC) resonances that occur at higher energies 
corresponding to photoexcitation of dipole transitions in the residual ion
by the incident photon (Yu \& Seaton 1987).
The PECs are very large features that attenuate the background cross
section by up to orders of magnitude, and are present in many
of the cross sections  in all ions with
target states coupled by dipole transitions, i.e.,  
even and odd parity LS terms. Much of the effort in the
decade-long Opacity Project was devoted to a careful consideration and
delineation of resonances using the R-matrix method  which has the
advantage that once the (e + ion) Hamiltonian has been
diagonalized in the R-matrix basis, cross sections may be computed at an
arbitrary number of energies to study resonant phenomena.
The methods and a number of calculations are described in the  volume
{\it The Opacity Project} by the Opacity Project Team (1995).

Owing to the complexity in the structure of the cross sections, 
thousands of points are normally calculated
to represent the detailed cross sections for each bound state of an ion
or atom. While this is a great advance in terms of 
atomic physics and accuracy, the huge amount and the inherent details of the
data do present a serious practical problem for numerical modeling.
An additional difficulty in the use of these cross sections for 
photoionization modeling is uncertainty in the precise positions of these resonances.
The OP cross sections were calculated primarily
for the computation of Rosseland and Planck mean opacities in 
local thermodynamic
equilibrium (LTE) for stellar envelope models (Seaton \etal 1994);
the cross sections at all photon frequencies are integrated over the 
Planckian black-body radiation field.
Whereas 
the LTE mean opacities are insensitive to small uncertainties in the
precise locations of resonances in photoionization cross sections, the
situation is different for the calculation of photoionization
rates of individual atomic species in non-LTE astrophysical 
sources photoionized by radiation fields that include spectral 
lines, such as H~II regions or active galactic nuclei (AGN). There may
be spurious coincidences between the strong lines and the narrow resonance
features in the original data. On the other 
hand, the physical presence of extensive structures of 
resonances in cross sections has a pronounced effect on the total
photoionization rate 
that should not be neglected, as exemplified  by
recent work on Fe ions (Nahar, Bautista, \& Pradhan 1997a, 
Bautista \& Pradhan 1998).  Thus, a numerical procedure is needed that
accurately and efficiently reproduces the new CC cross sections
for astrophysical modeling, in particular the extensive OP photoionization
cross section data in TOPbase.

 In some previous works analytic fits have been presented for
partial photoionization cross sections of sub-shells (Verner \etal 1993,
Verner and Yakovlev 1995), and for the OP 
cross sections ``smoothed" over resonances (Verner \etal 1996). 
However, analytic fits can not reproduce the well localized effects of
resonances and groups of resonances.
In their approach, Verner \etal
(1996) smoothed over resonances at variable energy 
intervals whose widths were adjusted until the resonance 
structures disappeared. In some cases, however, very large resonances 
could not be smoothed, so they were neglected in the fits. The analytic 
fits of Verner \etal seem computationally efficient for modeling 
computer codes. 
However, the smoothing procedure is unphysical and
artificially deletes most of the extensive resonance structures in the 
OP cross sections.
In addition, as we show later, neglecting 
very large resonances results in errors in the fitted cross sections 
and in the resulting photoionization rates. 
Although such errors in the photoionization rates are difficult to quantify 
in general, due to the frequency dependence of the 
irregular radiation field that varies 
from object to object and from one point to another within the same object,
we present several quantitative estimates for specific cases. 

In this paper we compute resonance-averaged photoionization  (RAP)
cross sections from a convolution 
with a running Gaussian distribution over energy intervals that subsume
the uncertainties in resonance positions, estimated to be about
1\% from a comparison of the calculated bound state energy levels
with spectroscopic measurements ({\em The Opacity Project Team}, 1995).
This procedure should minimize errors in ionization rates
due to inaccuracies in resonance positions,
while taking into account their contributions and preserving 
the overall physical complexities in the structure of the cross sections, 
especially in the
important near threshold region. Further, RAP cross sections effectively
simulate some broadening processes, notably thermal (Doppler)
broadening, that result in a natural smearing of the sharp resonance 
features. Thus, RAP cross sections assume a qualitatively physical form
even though the quantitative aspects may not be generalized for all 
sources.

The differences between photoionization rates calculated 
with the present RAP cross sections, and the detailed cross sections, are studied for 
a variety of radiation fields. 
The relatively low-energy cross sections from the OP and IP are 
merged with cross sections from Relativistic Distorted Wave 
calculations by Zhang (1997) for high photon energies including 
inner-shell ionization from closed sub-shells (not considered in the OP
data), and from the         
Hartree-Slater central-field calculations by Reilman \& Manson (1979). 
Further, we employ a numerical technique for 
representing the photoionization cross sections by a small number of 
points, from the photoionization threshold to very high energies. 
The tabulated cross sections can be readily coded in computer 
modeling programs to enable accurate computation of photoionization rates
for an arbitrary ionizing radiation flux.  A Fortran subroutine RESPHOT 
is made available to users to facilitate the interface of RAP data 
with models.

\section{Resonance-Averaged Photoionization (RAP) Cross Sections}
 
The uncertainty in the position of any given feature 
in the photoionization cross section may be represented by a probabilistic 
Gaussian distribution of width $\delta E$ around the position predicted 
by the theoretical calculation. There is, in principle, no reason to 
expect that the accuracy 
should vary with the central energy of a given feature or from one 
cross section to another. Thus we can 
assume $\delta E/E \equiv \Delta$ to be constant. 
Then the averaged photoionization cross section in terms of the detailed 
theoretical cross section convolved over
the probabilistic distribution is
\begin{equation}
\sigma_A(E)= C \int_{E_0}^{\infty}\sigma(x)
\,\exp{[-(x-E)^2/2(\delta E)^2 ]} \,\,dx,
\end{equation}
where $\sigma$ and $\sigma_A$ are the detailed and averaged photoionization 
cross sections respectively, E$_o$ is the ionization threshold energy, 
and $C$ is a normalizing constant.

Fig. 1 compares the detailed and the RAP cross sections 
for Fe~II (Nahar \& Pradhan 1994) 
and Fe~I (Bautista 1997) for choices of the $\Delta$ 
= 0.01, 0.03, 0.05, and 0.10. The convolved cross 
sections with the Gaussian distribution are smooth even across regions of 
intricate resonance structures. In regions free of resonances 
the RAP cross sections asymptotically approach the original cross sections 
without loss of accuracy. It is observed that the choice of the width 
of the Gaussian distribution has an appreciable effect on the resulting 
RAPs. As the width of the distribution decreases the 
RAPs show more structure resembling the detailed resonant structure.
On the other hand, if the chosen width of the distribution 
is too large the overall structure of the cross sections in the resonant 
region is smeared over and the background may be unduly altered.
This is seen in the case of Fe~II (Fig. 
1(a)) 
and Fe~I (Fig.1(b)) in the 0.8 to 0.9 Ry region that lies between 
large resonance structures. 
  
Based on such numerical tests, we adopt a standard width for all the cross 
sections of $\Delta=0.03$ (solid lines in Fig. 1(a) and (b)). 
This choice is sufficiently 
conservative with respect to the uncertainties in the theoretical cross 
sections and is able to provide RAP cross sections that resemble reasonably well
 the overall structure of the cross sections.
This choice of $\Delta$ also yields RAPs sufficiently smooth to be 
represented by small number of points that lead to accurate 
photoionization rates ((see Sections 4 and 5).

\section{The Data}

The OP ground state photoionization cross sections for 
atoms and ions of He through Si (Z = 2 -- 14), and S, Ar, Ca, 
and Fe are obtained 
from TOPbase (Cunto \etal 1993). For the lowest ionization stages of 
Iron, Fe~I -- V,    
radiative data of much higher accuracy than those from the OP have 
recently been computed under the IP (Table 1 of Bautista \& Pradhan
1998) and have been included in the
present work (detailed references are given in the Appendix). 
The R-matrix calculations performed under the OP and IP were carried out 
for photon energies up to just above  the highest target state in 
the CC expansion for the residual core
ion. The first version of TOPbase included OP cross sections with
power law tails extrapolated to energies higher than in the 
R-matrix calculations. 
High energy cross sections, however, have now been calculated for 
all the ground and excited states of 
atoms and ions with Z$\le 12$ using a fully relativistic 
distorted-wave method (Zhang 1997) that includes the inner-shell `edges'
not considered in the original OP data. The low energy R-matrix 
cross sections smoothly match the high energy distorted-wave 
tails; which yields a consistent set of merged (OP + RDW) results that 
should be accurate for all energies of practical interest. For the lowest 
ionization stages of S, Ar, Ca, and Fe that are not included in 
Zhang's calculations, we have adopted central-field 
high energy cross sections by Reilman \& Manson (1979).

\section{Numerical Representation of the RAP Cross Sections}
\label{RAP}

Having calculated the RAP cross sections, we represent these with
a minimum number of points selected so  
that the cross section can be recovered by linear interpolation
to an accuracy better than 3\%. 
We obtain a representation for the cross sections, 
from the ionization threshold to very high photon energies including 
all of the inner shell ionization edges, with approximately 30 points 
per cross section. Examples of
the RAP cross sections, and differences with the analytic fits (Verner
\etal 1996), are 
illustrated in Fig. 2 for S~I and Fe~I. It is clear that for these two
important elements these differences are substantial and would correspondingly
affect the photoionization rates. In particular it may be noted that
the effect of resonances varies significantly with energy, representative
of the complex atomic effects such as the Rydberg series limits that can
not be reproduced by any analytic procedure. For example, the resonance-
averaged structure in the RAP cross section for
S~I in the near-threshold region is a rise and a dip, corresponding to
actual resonances. The resonances make an even greater contribution for
Fe~I and the analytic fit is likely to yield a serious underestimation
of the photoionization of neutral iron.

Fig. 3 presents RAP cross sections for several other elements, with
some singularly  large features over wide energy ranges (e.g. Fe~IV and
Al~I). Keeping in mind
the 1\% or so uncertainty in the resonance positions, the RAPs represent
these physical features (an extensive discussion of the resonant feature
in Fe~IV is given by Bautista \& Pradhan 1997).
Fig. 3 also show the discrete sets of points that can be interpolated 
for a detailed and accurate representation of the averaged cross
sections. Such a representation is not possible with analytic fits. 
In addition to a more accurate representation of the atomic physics
the present RAP cross sections should also be computationally preferable to
analytic fits        
since a single set of points can reproduce the effective cross section, 
while analytic fits require several formulae and parameters 
for all of the inner-shell contributions.

\section{Accuracy of RAP Cross Sections}

A careful study of the accuracy of the RAP cross sections and their reduced representation 
is important if they are to be used for practical applications.   
Any reasonable transformation or smoothing procedure of the photoionization
cross sections should conserve the total area under the cross section function integrated over a
certain energy integral. This, however, gives no indication about the
actual accuracy of the transformed function.
A good indication of the uncertainty in the cross sections may be  obtained 
from the photoionization rates that result from the product 
of the cross section and 
a radiation field integrated  over the photon energy from the 
ionization threshold to infinity. 
The radiation fields in practice are complicated functions of frequency 
and may even vary from one point to another 
within the same object, as well as from object to object. Then, different 
radiation field functions would sample preferentially 
distinct energy intervals in the 
cross sections and may be used as an accuracy indicator.

For the present work we have selected nine different ionizing radiation fields 
that are expected to represent some general conditions 
for a number of cases 
of astrophysical interest. These radiation fields correspond to 
ionizing sources 
typical of an O star with \teff = 40,000 K, a high luminosity star with 
\teff = 100,000 K, and an extremely hot $T=10^8$ K black body source. 
Each of these sources is assumed to be surrounded by a gaseous envelope
with nearly 
cosmic chemical composition under pressure equilibrium conditions. The 
densities of the ionized gas at the near side to the source 
are taken to be $10^4,\ 3\times 10^3,$ and $10^{10}$ cm$^{-3}$, 
respectively. Under photoionization equilibrium the physical
parameters  were calculated 
for each of these nebulae using the computer code CLOUDY (Ferland 1993), 
and three different ionizing radiation fields were obtained for each case 
corresponding to the conditions at the near side of the cloud, 
at half depth of the ionized cloud, and near the ionization front.  
As an example, the radiation fields selected for the 
\teff = 100,000 K source are shown in Fig. 4. 

All detailed OP cross sections, their RAP cross sections, and their
reduced representations (linearly interpolated values between the
RAP cross sections), were integrated over the different radiation 
fields to obtain the photoionization rates and the results compared.
Photoionization rates obtained 
with detailed cross sections and RAP cross sections  agree within 
5\%. 

When comparing the photoionization rates with those obtained from
the analytic fits of Verner \etal (1996), 
significant differences are found.
The most prominent differences are for the lower ionization stages of iron 
Fe~I~--~V, and Na~VII, for which the fits of Verner \etal 
give photoionization rates 
differing by up to about 70\%. Differences between 20-30\% are found 
for Be~I, S~IV, Ar~II, Fe~VII, and Fe~XI, and between 10-20\% 
for B~I, S~VII, Mg~I, Mg~II, Al~I, Ar~I, Ar~III, Ar~V, and Fe~VIII. 
For all other 
ions the fits of Verner \etal yield photoionization rates that 
agree to within 10\% with the present results. This agreement for the
last set of data is primarily for the multiply ionized systems where the
resonances are usually narrow and the
resonance contributions, relative to  the background, are small.
It is emphasized that the errors in the photoionization rates when 
using analytic fits for the cross sections vary with the shape of the 
radiation field and are unpredictable in general.

\section{Recombination Rates and Ionization Fractions}

 A further check on the new RAP cross sections may be made by computing ionization
fractions in photoionization equilibrium. These calculations also
require (electron-ion) recombination rate coefficients. In recent years
a unified method has been developed that incorporates radiative and
dielectronic recombination (RR and DR) in an {\em ab initio} manner, and
enables the calculation of total (e + ion) recombination rates in the CC
approximation using the R-matrix method (Nahar \& Pradhan
1992, 1995). In addition, the new recombination rates  are fully
self-consistent with the photoionization cross sections as both the
photoionization and the recombination data are calculated in the CC
approximation using the same eigenfunction expansion over the 
states of the residual ion. Unified, total recombination rates have been
computed so far for approximately 33 atoms and ions, including all C, N, O
ions (Nahar and Pradhan 1997, Nahar 1998), the C-sequence ions (Nahar
1995,1996), and Fe ions Fe~I -- V (references are given in Bautista \& Pradhan 1998).

 In a recent  work on iron emission and ionization structure in gaseous
nebulae (Bautista \& Pradhan 1998) the new
photoionization/recombination data for Fe~I -- V, including detailed and
RAP cross sections, 
was employed to obtain
ionic fractions of Fe in a photoionized H~II region (the Orion nebula),
and considerable differences were found with previous works.
In this work we compute a few C, N, O ionization fractions
 using the RAP cross sections computed in
the present work and the new unified (e + ion) recombination rates, to
study the effect of the new photoionization/recombination data.
Although the differences for lighter elements are relatively smaller
than for the Fe ions,
they can be significant in temperature ranges in transition regions
between adjacent ionization stages, as illustrated in Fig. 5. Whereas
there is no significant difference in the RAP cross sections computed in
this work for C,N, and O, and the earlier ground state photoionization data incorporated
in CLOUDY, some significant differences are found when the new unified
recombination data for C,N,O ions is employed.

\section{FORTRAN Subroutines and Data}  
                     
The  RAP cross sections for the ground state of all atoms and ions in
TOPbase, and new data for Fe~I~--~V and some other ions, are available in
a FORTRAN subroutine RESPHOT that can be readily interfaced with
modeling codes. The routine is available 
electronically from the authors. Given an ion stage (N,Z) and an energy in 
Rydbergs, RESPHOT returns the linearly interpolated value 
of the photoionization cross section from the table of points 
described in Section 4. It is also important to point out that the 
RAP cross sections are given as a function of the energy of the 
ejected electron (i.e., energies with respect to the ionization 
threshold) instead of the photon energies in TOPbase that are relative
to the first ionization potential calculated to about 1\% accuracy.
Users can easily scale the 
cross sections to the more accurate experimental ionization
potentials.

Also available is a routine RCRATE which gives total 
unified e-ion recombination rates that
may be used instead of the combination of previously available (and
often inaccurate) data on RR and DR rates. 
Given an ion stage (N,Z) and an electron temperature, RCRATE returns the total
recombination rate coefficient interpolated from the tables 
of Nahar (1995,1996), Nahar and Pradhan (1997), and the references
in Table 1 of Bautista and Pradhan (1998).

RESPHOT and RCRATE can be easily interfaced with photoionization codes 
such CLOUDY, as demonstrated by their use to produce the results in Figs.
4 and 5.

\section{Conclusion}
                     Resonance-averaged photoionization (RAP) cross sections
have been calculated for most atoms and ions of astrophysical
importance using the Opacity Project data from TOPbase and new data on
Fe and C,N,O ions. 
These incorporate the effect of autoionizing resonances in
an averaged manner that is not too sensitive to the precise positions of
resonances, but accounts for the often significant attenuation of the
effective cross sections that is neglected in earlier works. The
RAP cross sections have been represented by a small number of points that can be
readily interpolated in modeling codes to reproduce photoionization
cross sections at all energies of practical interest, including
inner-shell ionization thresholds. Illustrative examples show 
considerable differences with analytic fits that neglect resonance
structures. It is also pointed out that new recombination rates, 
unifying the radiative and dielectronic recombination processes, are
being computed for astrophysically abundant elements to provide a
self-consistent set of photoionization/recombination data for
modeling astrophysical sources in radiative equilibrium.

\acknowledgements
                   This work was partially supported by a NSF grant for the
Iron Project PHY-9482198 and the NASA Astrophysical Data Program. PR acknowledges the financial support from the Graduate School at OSU through a University Fellowship.

\section*{Appendix}

\begin{deluxetable}{ll}
\small
\singlespace
\footnotesize
\tablewidth{30pc}
\tablecaption{References to the Photoionization Cross Sections}
\tablehead{
\colhead{\bf Ion} &
\colhead{\bf Reference} }
\startdata
He-like & Fernley \etal 1987 \nl
Li-like & Peach \etal 1988 \nl
Be-like & Tully \etal 1990\nl
 B I    & Berrington \& Hibbert, in preparation \nl
B-like  & Fernley et al., in preparation \nl
C-like  & Luo \& Pradhan 1989 \nl
N-like  & Burke \& Lennon, in preparation \nl
O-like, F-like & Butler \& Zeippen, in preparation \nl
Ne-like & Scott, in preparation \nl  
Na-like & Taylor, in preparation \nl
Mg-like & Butler \etal 1993 \nl
Al-like & Mendoza \etal 1995 \nl
Si-like & Nahar \& Pradhan 1993 \nl
P-like  & Butler et al., in preparation \nl
S-like  & Berrington et al., in preparation\nl
Cl-like & Storey \& Taylor, in preparation\nl
 Fe~IX  & Butler et al., in preparation \nl
Ar-like & Saraph \& Storey, in preparation 
\tablebreak
 Fe~VIII& Butler \etal, in preparation \nl
 Fe~VII & Sawey \&  Berrington 1992 \nl
Ca-like & Saraph \& Storey, in preparation \nl
 Fe~VI  & Butler et al., in preparation \nl
 Fe~V   & Bautista 1996 \nl
 Fe~IV  & Bautista \& Pradhan 1997 \nl
 Fe~III & Nahar 1996 \nl
 Fe~II  & Nahar \& Pradhan 1994 \nl
 Fe~I   & Bautista 1997 \nl
\enddata
\end{deluxetable}


\def \aas {\it Astronomy and Astrophys. Supplement Series}
\def \rmex {Rev. Mexicana de Astronom\'{\i}a y Astrofis.}

 
\clearpage

\begin{figure}
	\null
	\epsfxsize=15truecm
	\epsfysize=15truecm
	\hspace{+0.5truecm}
 	\epsfbox{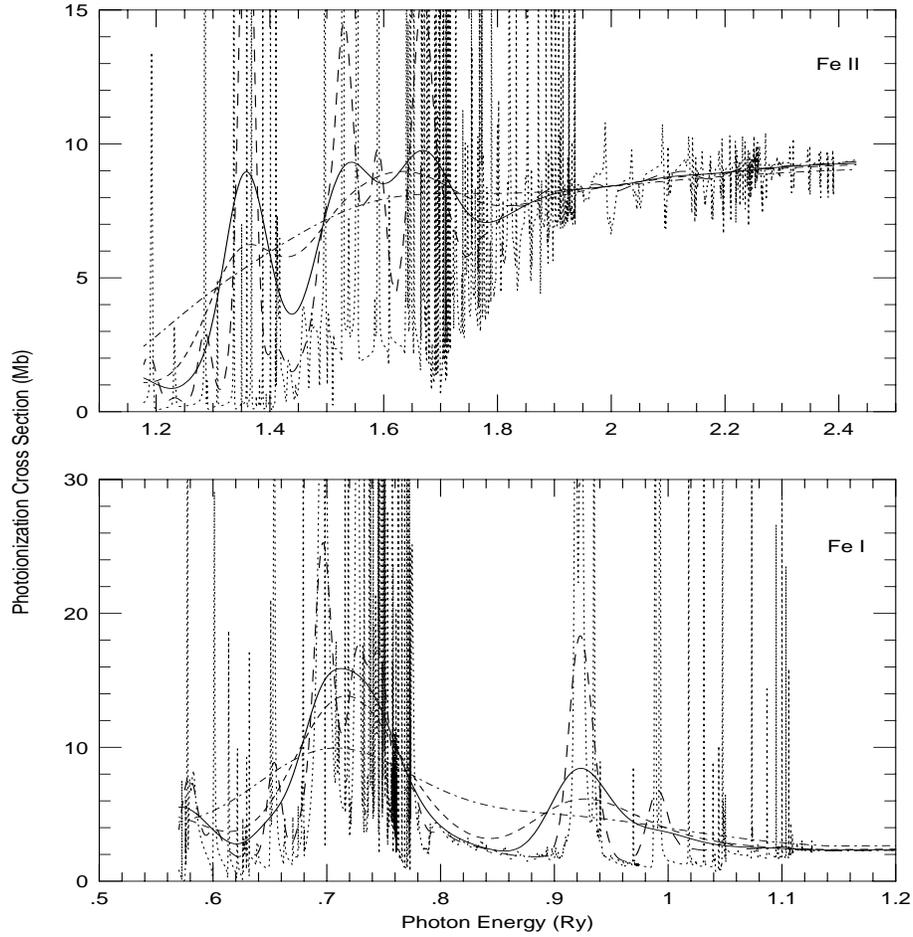}
\caption{
Ground state photoionization cross sections 
for Fe$^+$ (a) and Fe$^0$ (b). The detailed cross sections are shown 
by the dotted line and the resonance-averaged cross sections are 
shown for $\delta E/E = 0.01$ (long dashed line), 0.03 (solid line);
0.05 (short dashed line), and 0.10 (dot-dashed line). }
\end{figure}

\clearpage	

\begin{figure}
	\null
	\epsfxsize=15truecm
	\epsfysize=15truecm
	\hspace{+0.5truecm}
 	\epsfbox{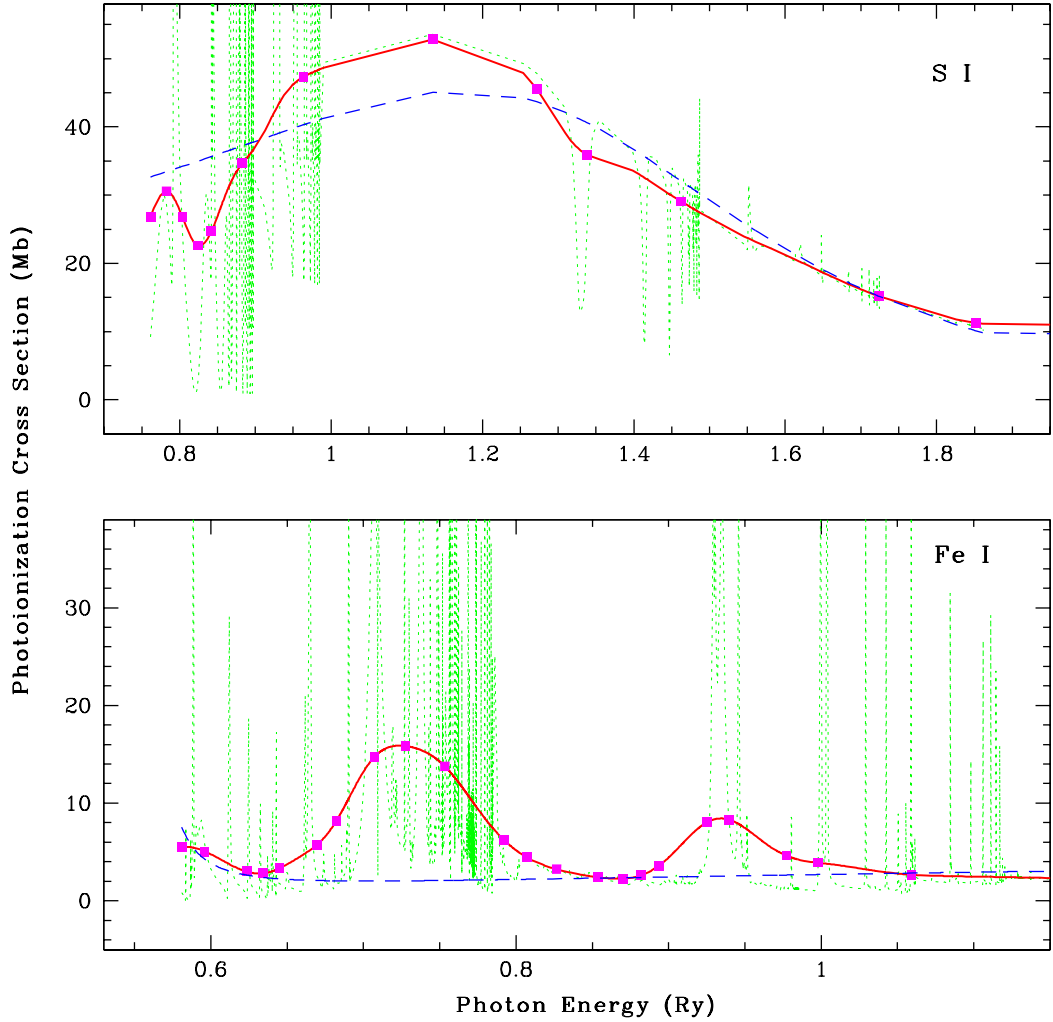}
\caption{Sample of resonance-averaged photoionization
(RAP) cross sections (solid line) for S~I and Fe~I; 
filled squares are points chosen to represent the RAP cross sections. 
The short-dashed line is the detailed OP cross section, 
the long-dashed line is the cross section from analytic fits by 
Verner et al. (1996).}
\end{figure}

\clearpage	
 
\begin{figure}
	\null
	\epsfxsize=15truecm
	\epsfysize=15truecm
	\hspace{+0.5truecm}
 	\epsfbox{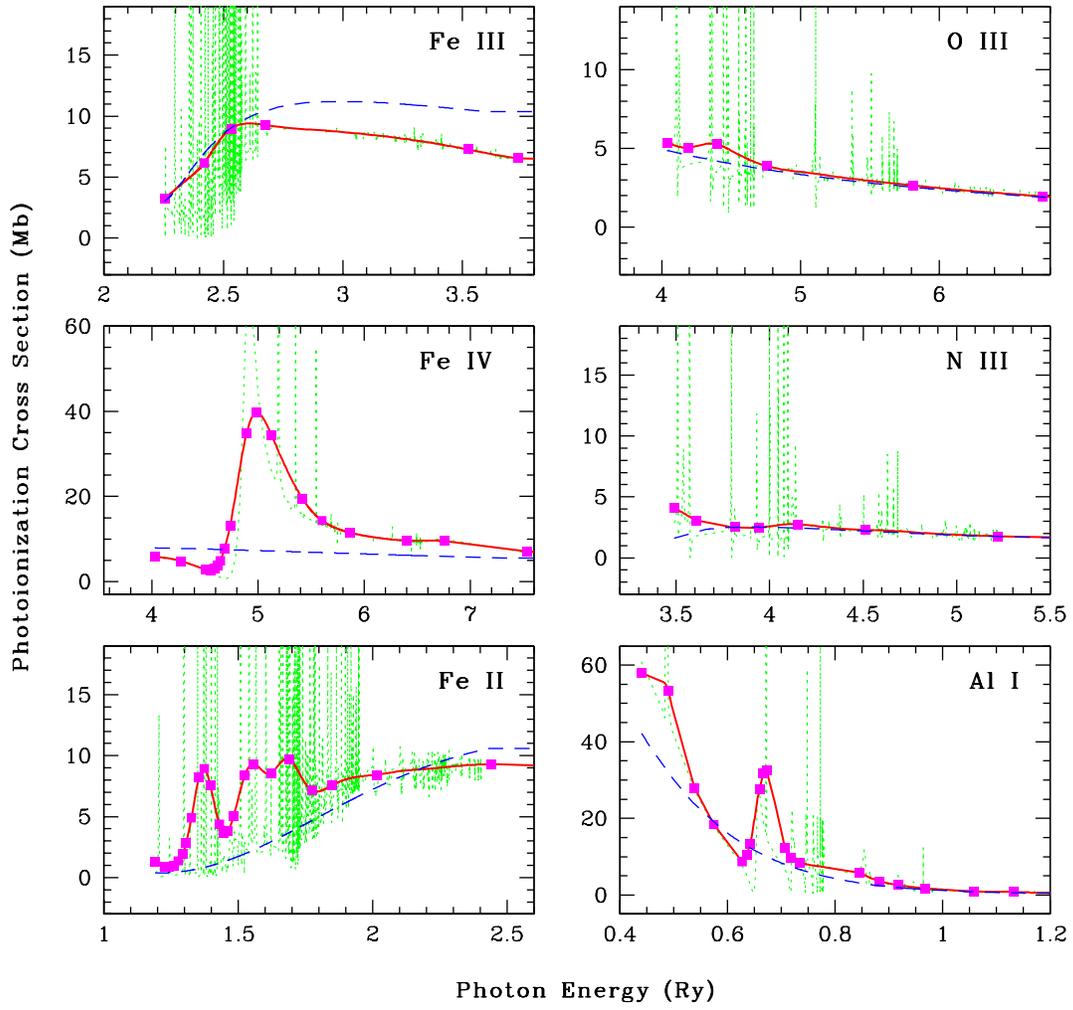}
\caption{Sample of RAP cross sections for several other
ions (as in Fig. 2).}
\end{figure}

\clearpage	

 \begin{figure}
	\null
	\epsfxsize=15truecm
	\epsfysize=15truecm
	\hspace{+0.5truecm}
 	\epsfbox{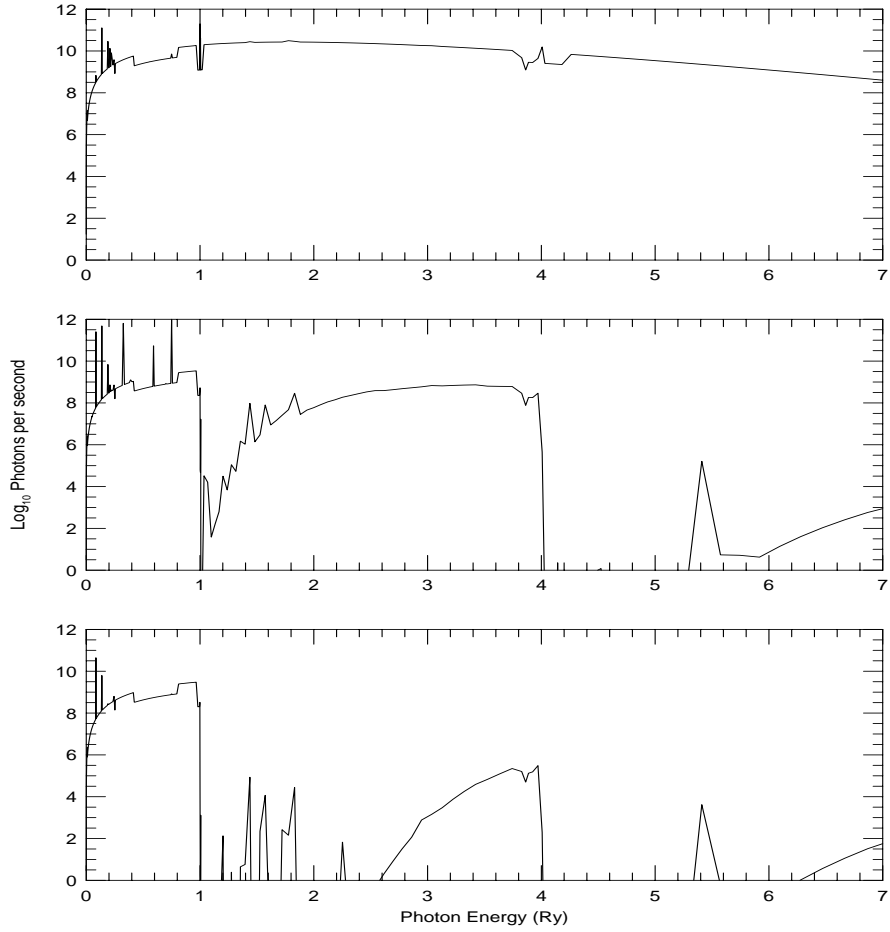}
\caption{
Sample of ionizing radiation fields at different 
points in a photoionized nebula. 
The radiation fields shown are those at the 
near side of the cloud, at half depth,  and near the ionization front, 
(top to bottom respectively) for a star of effective temperature of 
$10^5$ K.
}
\end{figure}

\clearpage	

\begin{figure}
	\null
	\epsfxsize=13truecm
	\epsfysize=13truecm
	\hspace{+0.5truecm}
 	\epsfbox{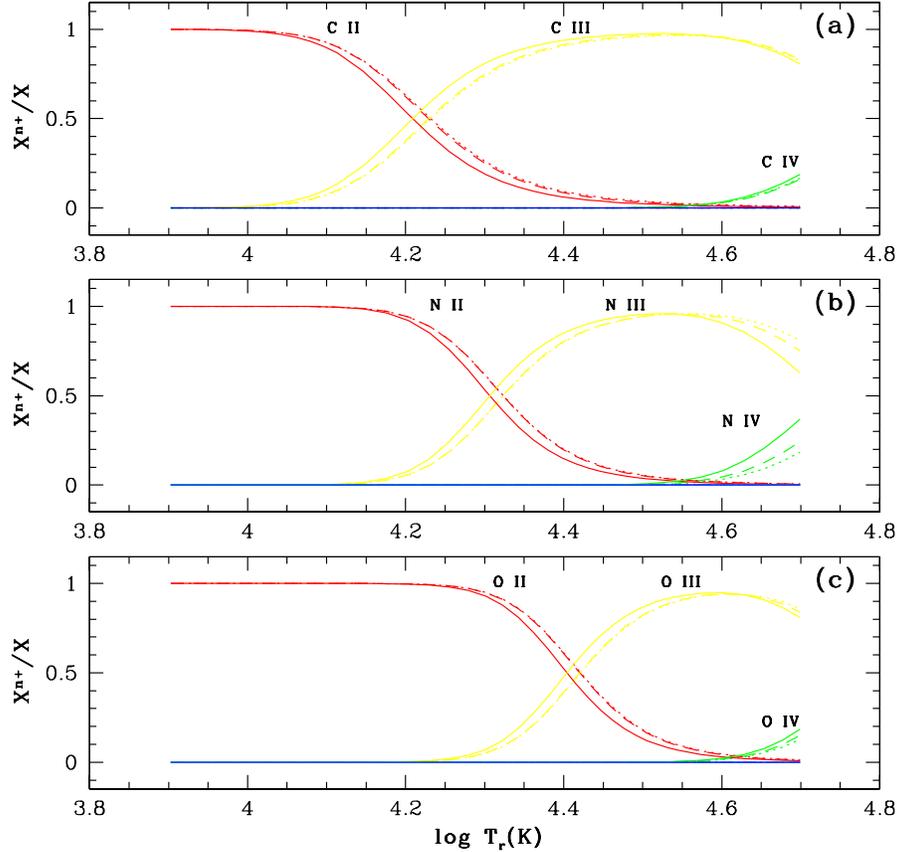}
\caption{
Ionization fractions for Carbon, Nitrogen, and Oxygen ions 
as a function of the radiation temperature $T_r$ of the black body 
ionizing source: with new RAP cross sections from subroutine RESPHOT, 
and new unified recombination rates from subroutine RCRATE (solid lines), 
with new  RAP cross sections and earlier recombination data as in CLOUDY
(dashed lines), and with phototionization and recombination
data as in CLOUDY (dotted lines); the calculations are 
for photon flux per cloud unit area of 10$^{13}$ cm$^{-2}$, 
H density of 10$^4$ cm$^{-3}$, and electron temperature of 10$^4$ K.
The curves show that the new unified recombination data for C,N,O leads 
to some
significant differences in the ionization fractions.}
\end{figure}

\clearpage

\end{document}